\documentclass[12pt, letterpaper]{article}
\usepackage[utf8]{inputenc}
\usepackage{mathtools}
\usepackage{amssymb}
\usepackage{physics}
\usepackage{setspace}
\usepackage{cite}
\usepackage{xcolor}
\begin{document}

\begin{center}
\textbf{\Large{Critical review of prevailing explanations for the emergence of classicality in cosmology}}
\end{center} 

\begin{center}
\large{Javier Berjon,$^{1}$ Elias Okon$^{1}$ and Daniel Sudarsky.$^{2}$}\\
\end{center}

\begin{center}
$^{1}$Instituto de Investigaciones Filos\'{o}ficas, Universidad Nacional Aut\'{o}noma de M\'{e}xico, Mexico City, Mexico.\\
$^{2}$Instituto de Ciencias Nucleares, Universidad Nacional Aut\'{o}noma de M\'{e}xico, Mexico City, Mexico.\\
 \end{center}

\begin{abstract}
There have been recent attempts at justifying, from first principles, and within the standard framework, the emergence of classical behavior in the post-inflationary cosmological context. Accounting for this emergence is an important issue, as it underlies the extraordinary empirical success of our current understanding of cosmology. In this work, we offer a critique of different efforts at explaining the emergence of classical behavior in cosmology within the standard framework. We argue that such endeavors are generically found lacking in conceptual clarity, as they invariably rely, either upon unjustified, implicit assumptions, or on circular logic. We conclude that, within the standard approach, the emergence of classical behavior in cosmology constitutes an unexplained phenomenon.
\end{abstract}

\onehalfspacing
\section{Introduction}

Contemporary cosmology tries to explain the emergence of structure in the universe. The current explanation for such an emergence is based on the study of certain quantum effects, or \emph{fluctuations}, during the inflationary epoch. The starting point of the standard account is a completely homogeneous and isotropic situation---both as far as space-time and matter are concerned---and quantum effects over such a symmetric background are supposed to break the symmetry. That is, these quantum effects are assumed to source the primordial inhomogeneities and anisotropies that, according to the theory, eventually evolve into all the structure we observe in the late universe---including the formation of galaxies, stars and planets (of which, in at least one, intelligent life capable of wondering about its own origins developed). The primordial seeds responsible for the formation of structure are believed to leave their first observational imprint (insofar as current technology allows us to confirm) in the anisotropies of the cosmic microwave background (CMB). 

Based on these ideas regarding quantum fluctuations, what is usually done \emph{in practice} is the following. When considering the late inflationary epoch, it is argued that the quantum state (initially a Bunch-Davis vacuum) can be replaced with a \emph{classical description} of the state of affairs. More specifically, at the end of inflation, one usually disregards the original quantum nature of the system and, instead, describes it as an element of an ensemble of systems, characterized by a certain distribution of states in classical phase space. Cosmologists have tried to account for this \emph{quantum-to-classical transition}, or \emph{emergence of classical behavior}, by different types of arguments, and relying on a variety of methods. In a recent paper by Ashtekar et al., \cite{ashtekar}, some of these explanations for the emergence of classicality are revisited. In particular, they explore: i) the fading of the importance of quantum non-commutativity, ii) the phenomenon of quantum squeezing, and iii) the ability to approximate the quantum state by a distribution function on the classical phase space. The conclusion reached in \cite{ashtekar} is that, while the three notions considered are conceptually distinct, they do provide a rather robust explanation for the emergence of classicality in cosmology---and do so even in more general contexts.

In this work, we challenge the notion that these or similar schemes constitute a solid justification for the transition to a classical description of the situation under consideration. In our opinion, all of the standard attempts simply fail. In particular, we find them lacking in conceptual clarity and invariably relying upon unjustified, implicit assumptions. We conclude that the standard approach to cosmology is simply unable to explain the emergence of classical behavior.

To make our case, we start in section \ref{Class} by considering in full generality the issue of \emph{classicalization of a quantum system}. In particular, we consider the ways in which such a notion might be defined, and the conditions under which its use could be deemed appropriate. Next, in section \ref{ClassCos} we explore the issue of classicalization, as it arises in the cosmological context in particular. Then, in section \ref{Ash} we recount the arguments and conclusions in \cite{ashtekar} and we examine whether they succeed in the job they set out to do. We finish the discussion with some concluding remarks in section \ref{Conc}. 

In this arXiv version of the paper, we include as Section \ref{reply} our reply to a note added in arXiv:2004.10684 in response to an earlier version of this paper.

\section{On the classicalization of a quantum system}
\label{Class}

We start the discussion by emphasizing that an exploration of the idea of the classicalization of a quantum system must develop in a setting in which one takes the quantum description to be, if not fundamental, at least deeper or more fundamental than the classical one.\footnote{We assume this to be the case for all types of matter, an even for space-time itself. There are, of course, arguments contemplating the viability of a theory in which space-time is in fact classical, without any quantum theory of gravity underlying it (see, e.g., \cite{Carlip}). Those considerations have raised quite interesting debates, but we will not contemplate such possibilities any further in the present work.} If so, it is clear that there are no grounds for interpreting \emph{classicalization} in a literal sense; in other words, it is never the case that a quantum system, strictly speaking, stops being quantum and becomes `fully classical'. So what is actually meant when such a concept is employed? A sensible possibility is to take it to reflect situations in which certain aspects of the system in question can be suitably described through the use of classical language. This sounds reasonable, but rather vague, so, in order to clarify the idea, we need to spell out the notions sketched in the previous statement.

To do that, we consider first the relatively simple situation for which quantum theory was initially conceived, namely, the use of the theory to analyze a small system in the lab. For such a situation, we can use the following \emph{instrumentalist} criteria for classicalization:
\begin{quote}
A quantum system described by the state $|\psi \rangle$ is said to have \emph{classicalized} during the time interval $T$, relative to the measuring apparatuses $\lbrace M_i \rbrace$ designed to measure the observables $\lbrace \hat O_i \rbrace$, if during that interval:
\begin{description}
\item[\normalfont{i)}] The quantum uncertainties of $|\psi \rangle$ regarding the observables $\lbrace \hat O_i \rbrace$ are equal or smaller than the experimental uncertainties dictated by the $\lbrace M_i \rbrace$.
\item[\normalfont{ii)}] To the level of accuracy provided by the $\lbrace M_i \rbrace$, the expectation values of the observables $\lbrace \hat O_i \rbrace$ approximate the corresponding classical equations of motion.
\end{description}
\end{quote}
The first condition ensures that the results of the available measurements coincide with the corresponding expectation values. It also ensures that the disturbances produced by such measurements are negligible, so they can be safely ignored. The second condition guarantees that, within the accuracy of the available measurements, the expectation values follow the corresponding classical equations of motion---and hence that the measured values agree with the classical predictions.

As an example, consider a simple harmonic oscillator in a highly excited coherent state, provided with apparatuses to measure position and momentum with sufficiently low accuracy. Such a system clearly fits our classicality criterion. In contrast, consider the same harmonic oscillator, this time in its ground state $ |\psi_0 \rangle$, together with a position measuring device of accuracy greater than $ \Delta \hat X _{ |\psi_0 \rangle}$. In that case, measurements of position would, in general, not coincide with the expectation value of position and, moreover, they would radically change the state of the system. Clearly, such a situation does not fit our classicality criterion. Suppose, however, that after a position measurement, the system ends up in a coherent state and that, after the measurement, the accuracy of the measuring apparatus is lowered to below the width of the resulting coherent state. If so, one ends up with a system for which the classicality criterion is valid. That is, one starts with a situation that does not fit the classicality criterion, but after a measurement, and provided that only particular measurements are allowed, a sort of classicalization arises.

Now, if we were to consider a large ensemble of harmonic oscillators, all initially prepared in the ground state, measurements of position would produce a large range of results, spread around the minimum of the potential (say, $X=0$), with dispersion $ \Delta \hat X _{ |\psi_0 \rangle} $. Moreover, if as in the example above, such measurements would lead to classicalization, then the ensemble could be effectively described by a distribution over classical states, characterized by positions distributed around the origin with a statistical spread equal to $ \Delta \hat X _{ |\psi_0 \rangle}$. It is important to point out that it is very unfortunate, due to the confusion it tends to generate, that both the statistical dispersion of the distribution of the ensemble after a measurement, and the quantum uncertainty or width of the original state, are usually referred to by the same word: \emph{fluctuations}. Let us emphasize that, even though they might have the same numerical value, conceptually, they are very different notions; and that, at least according to textbook quantum theory, their connection is only brought about through measurements.

A further aspect of the above example is worth considering, as it will strongly impinge on our coming discussion of the cosmological context. Having set the minimum of the harmonic oscillator potential at $ X=0$, it is clear that the dynamics of the system, as encoded in the Hamiltonian, has the reflection symmetry $ {\cal \ P}: X\to -X $. It is then natural to consider the behavior of states under such a symmetry, which can be associated with an operator ${\cal \hat P}$ with eigenvalues $+1$ (completely symmetric states) and $-1$ (completely anti-symmetric states). All states can be decomposed into their symmetric and anti-symmetric parts, with the vast majority presenting components of both kinds, and therefore having no well-defined behavior under the symmetry. Note, however, that the ground state is completely symmetric. Moreover, its unitary evolution will not break this symmetry (in fact, the symmetry will not be broken, even if we make the spring constant and mass time dependent, so as to lead to the squeezing of the state). Nevertheless, once a position measurement is carried out, the resulting state will, in general, be neither symmetric nor anti-symmetric: the symmetry of the initial state is then broken by the act of measuring $X$. In sum, the presence of `quantum fluctuations' in the ground state, i.e., the fact that the state has a non-zero width, in no way implies that it is not symmetric; it is only through the act of measuring $X$ that the symmetry is broken.

Before moving on, we discuss another illustrative example. Consider a free particle prepared in a state with a relatively small dispersion in momentum $\Delta P_0$ around the value $P_0$. Suppose also that the phases in the state are such that the uncertainty in position starts very large, then shrinks to a minimum $ \Delta X_{min } = \hbar / \Delta P_0$ at a certain time $T_{min} $, after which it spreads out again. Finally, suppose that an observer has measuring devices that work with precision $ \delta P$ and $ \delta X$ with $ \delta P > \Delta P_0 $ and $\delta X $ slightly larger than $ \Delta X_{min}$. One might then say that there is a certain time interval around $ T_{min} $ in which, according to our definition, the system classicalizes---both regarding measurements of $X$ and $P$ (note that the system behaves classically regarding $P$ at all times). What we want to emphasize is that, in this case, the classicalization is generated by the system's own dynamics, and not as a result of a measurement or external intervention, as it occurred in the previous example.

As far as we know, nothing of what we said above is new or controversial. Still, the discussion so far has provided us with some important lessons: 
\begin{enumerate}
\item Classicality can be predicated of a state of a system only when supplemented by a list of admissible observables, each associated with a tolerance in the uncertainty of its value. In an experimental setting, this list of observables and tolerances is dictated by the apparatuses available to the observer. As a consequence of this, there is no \textit{absolute} sense in which a quantum system might be said to classicalize---this can only be asserted \textit{relative} to the cited additional information. 
\item Classicalization, in the above described sense, might arise as a result of a measurement, or as a result of some other process (we provided examples for the two situations above). This, however, does not change the fact that the criterion for classicality we have been exploring crucially depends on the notion of measurement.
\end{enumerate}

The classicalization criterion introduced above was designed with standard laboratory situations in mind. It seems clear, though, that its use can be extended to more general situations, as long as the system under consideration is assumed to be describable by quantum theory, and as long as one is provided with a list of relevant observables, along with their tolerated uncertainties. It is important to point out that, in order to apply our proposed criterion for classicality, it is not necessary for actual measurements to ever take place; that is, even though, in order to define classicality, one must consider how a system would behave if measured, the system in question does not, in fact, have to be measured in order to fulfill the criterion. Therefore, a system can be deemed to have classicalized, even if there are no observers around to verify it. Still, the notion of measurement remains central for classicalization, as it does in the standard quantum interpretation. It seems clear then, that, whichever way we decide to define the notion of classicality, and as long as we retain the standard interpretation at the basis of the discussion, the notion of measurement will have to play a crucial role.

Regarding the distinct role that measurements play in standard quantum mechanics, it might be argued that such a role is to be expected because, ultimately, ``physics is just about measurements.'' That is, an argument could be made to the effect that it is only through measurements that we can interact with the world in order to compare it with our theoretical predictions---and that this is true, not only within quantum mechanics, but also classically. We should point out, though, that the notion of measurement enters quantum and classical theories in radically different fashions. While in classical theory the notion of measurement is relatively innocuous, in quantum theory, at least as depicted by the standard framework, a measurement enters the theory in an essential manner.

In classical theory, on the one hand, the notion of measurement does not appear at the foundational level, i.e., it is not part of its axiomatic structure. Sure, it might be argued that classical theories fail to precisely define what counts as a measurement, and even that the notion might be ultimately tied to the notion of perception, and even to the hard problem of consciousness \cite{chal}, but the point is that one does not need to know what constitutes a measurement in order to employ classical physics to make predictions. In other words, once the precise constitution and initial state of both a system and a measuring device are provided, the dynamics of the theory itself accounts for the way in which the device ends up registering the result of a measurement.

In standard quantum theory, on the other hand, the notion of measurement enters the theory at the foundational or axiomatic level, and does so as a primitive, undefined term. This makes measurement an essential component of the theory, without which no predictions at all can be extracted from the formalism. That is, even if complete information of the initial state of the system and measuring apparatus are provided, in the absence of external information regarding what exactly constitutes a measurement, the dynamics of the theory itself simply does not lead to a final state in which the device ends up registering the result of a measurement. This situation might be argued to be tolerable in a lab setting. However, the fact that the standard framework crucially depends on measurements becomes untenable as soon as one intends to employ the theory in more complex scenarios, such as the early universe, in which there are no observers around to measure, or even worse, in which one is precisely trying to explain the emergence of the conditions that make possible the appearance of observers capable of performing measurements. In sum, one should not confuse the lack of a detailed description of what constitutes a measurement in classical theory, with the fact that, in standard quantum mechanics, observers external to the studied system are necessary in order to make sense of the theory.

A final important difference between measurements in classical and quantum theories, which will play a fundamental role in the discussion below, is the fact that in the latter, but not the former, the act of measurement generically changes the state of the measured system. That is, while in classical physics one assumes that one can always perform measurements that do not modify the state of the system, in quantum theory one recognizes that, unless one is measuring a state which is an eigenstate of the measured property, measurements will certainly modify the state of the system. This implies that, in quantum theory, measurements cannot be interpreted as simply revealing preexisting features of the measured system---a fundamental fact that must be kept in mind while interpreting the application of quantum theory to the early universe, as we do next.

\section{Classicalization in a cosmological context}
\label{ClassCos}
 
As we explained above, the starting point of the standard, contemporary account of cosmology is a fully homogeneous and isotropic situation---both as far as space-time and matter fields are concerned.\footnote{Inflation is assumed to quickly smooth out any preexisting inhomogeneities and anisotropies.} What is at stake, then, is the ability to account for the emergence of the \emph{primordial} seeds of cosmic structure. That is, an explanation of how the primordial inhomogeneities and anisotropies that, according to the theory, eventually evolve into all the structure we observe in the late universe, came about. The standard story for such an emergence runs as follows: quantum fields are subject to uncertainty relations that lead to inevitable quantum fluctuations; such fluctuations, which cannot be switched off even in principle, allow for the Bunch-Davies vacuum\footnote{Or a suitable adiabatic vacuum if the cosmic expansion is not exactly de Sitter. This caveat applies to every mention of the Bunch-Davies vacuum but will not be repeated elsewhere.} to be replaced, towards the end of inflation, with a distribution function on the classical phase space---and for its subsequent evolution to be described in classical terms. The key question, of course, is whether this standard procedure is justified; that is, if it indeed follows from first principles or if it depends upon some additional, possibly unwarranted, assumptions. 
 
As we saw in the previous section, in order to talk about the classicalization of a quantum system, it is essential to provide a well-defined sense in which the term is to be used. In particular, we saw that classicalization can only be defined relative to a set of observables, with associated uncertainties, and that a given state of the system might be describable in classical terms when focusing on a certain feature, but not when focusing on others. Therefore, to explore the issue of classicalization in a cosmological context, the first thing we must do is to establish the precise notion of classicality that must be employed, together with the set of relevant observables that should play a role in the analysis. Only then, a thorough assessment of the issue can be carried out.

Regarding the set of relevant observables, focus is often directed towards the inhomogeneities and anisotropies that are left as an imprint of the evolved version of those primordial inhomogeneities and anisotropies on the last scattering surface--- a feature which is susceptible of empirical analysis through detailed observations of the CMB. It is true that these are quite interesting, as they provide us with the opportunity to study those imprints before the much more complex effects of gravitational clustering take place. However, we must not lose sight of the fact that, focusing on such observations might be convenient, but is not the essence of the question. In fact, it is only a contingent fact that we can observe them at all.\footnote{For instance, if our solar system was deeply embedded in the bulge region of a gigantic galaxy, the CMB would be hidden from us (as it occurs with that part of the last scattering surface that lies on our galactic plane). Moreover, if another advanced civilization arises in the distant future, many billions of years after the current epoch, they might encounter the CMB so highly red-shifted that it might become unobservable in practice even by their best technology.} The aspect that is truly central to the story that our cosmological models must provide, and part, in fact, of the essential reason for doing cosmology, is to account for the primordial generation of structure, as well as for the evolution in time of this essential process that makes our own existence possible.

A key point that follows from all this is that, in a cosmological context, the issue of classicalization is intimately intertwined with the breakdown of the homogeneity and isotropy of the initial state; but what is exactly the relation between the two? It seems clear that the breakdown of the symmetry can occur in the absence of classicalization. That is, the mere breakdown of the symmetry does not imply classicalization (there are, of course, plenty of inhomogeneous and anisotropic quantum states that are far from satisfying the conditions for classicality). Therefore, in order to explain the emergence of classicality in the cosmological context, one must not only account for the breakdown of the symmetry, one also has to show that the resulting state allows for a classical description. 

Now, in the same way that a breakdown of homogeneity and isotropy does not imply classicality, classicality can occur in the absence of a breakdown of the symmetry. Nevertheless, what we want to point out is that, classicalization by itself, cannot erase or eliminate a symmetry present at the quantum level and, in particular, it cannot break homogeneity and isotropy. This, in fact, follows from very general considerations regarding a hierarchical structure of relations between theoretical descriptions. Suppose that we have, for the same system or class of systems, two theoretical frameworks: theory A, the more fundamental description, and theory B, an effective construction which, at least in principle, is derivable from theory A under a certain set of conditions or approximations. In such a situation, it is said that theory B supervenes on theory A, in the sense that differences according to theory B require differences according to theory A, \cite{Super}. If so, it follows that, if according to theory A, a physical situation possesses a certain symmetry, then the same physical situation, as described by theory B, also must possess the symmetry (provided that the symmetry in question is definable in terms of theory B).

As an illustration of this, consider a classical gas, described in terms of two different coarse-grainings, A and B, with the scale of the coarse-graining of A smaller than that of B. It is then quite possible for the characterization of the gas to be homogeneous according to B, but not to A. However, it is logically impossible for it to be homogeneous according to A, but not to B. Similarly, for the cosmological scenario at hand, if the quantum state of the system is indeed fully homogeneous and isotropic, as it is taken to be after the first few e-folds of inflation, then, even if a classical description would be available, such a description would also have to be fully homogeneous and isotropic. That is, classicalization by itself is unable to describe the emergence of structure. From this, we conclude that \emph{any successful effort to explain the emergence of classical behavior in the standard cosmological context must necessarily involve a satisfactory account of the breakdown of the homogeneity and isotropy of the quantum state}. But what constitutes a satisfactory account of the breakdown of such a symmetry?

The first thing to point out is that a purely unitary evolution of the quantum state after inflation is simply unable to break it. That is, the standard quantum evolution explicitly preserves the homogeneity and isotropy of the initial state---and crucially, this is so even if one takes into account the alleged factor responsible for the quantum-to-classical transition, namely quantum fluctuations. To see this, we note that such quantum fluctuations are no more than a way to express the fact that, regarding certain observables, the Bunch-Davies vacuum has some width and, in the same way that the fact that the ground state of a harmonic oscillator is symmetric regarding a reflection about the center of the potential, even though it has some width, the Bunch-Davies vacuum is fully homogeneous and isotropic, even though it is not sharp regarding certain variables. Moreover, it is clear that decoherence is also of no use in this respect because, a mere anthropocentric decision to call certain degrees of freedom the `system', and others the `environment', does not cancel the fact that the dynamics is fully unitary and the state fully symmetric.

It must be noted that there is another way in which the notion of decoherence could be understood---a notion that has been argued to be more relevant in the cosmological context at hand. Instead of linking decoherence to the tracing of environmental degrees of freedom, it could be associated with the wave function splitting into branches of disjoint support on configuration space (these branches, by the way, are what many Everettians would read as ``different worlds’’). The point we would like to stress is that this alternative notion of decoherence is of no use in order to explain the breakdown of the homogeneity and isotropy. This is because, since the wave functions associated with the quantum vacuum are Gaussian, there is no splitting of this sort ever occurring.

What about measurements? Well, it seems clear that any account of the breakdown of such a symmetry cannot rely on the notion of measurement since, clearly, in the homogeneous and isotropic state of the early universe, there cannot be observers to perform any measurements. One could, instead, claim that it is \emph{our} observations that break the symmetry, but an account of that sort clearly contains a circular causal chain of events: the breakdown of the homogeneity and isotropy of the early universe occurred because measurements were performed, but such measurements were possible because the symmetry was broken. That is, we are here because the seeds of structure that appeared in the early universe evolved into large scale cosmic structure (including galaxies and solar systems), but those seeds of structure arose because we performed measurements of the CMB. In short, our observations would be part of the account of the emergence of the conditions that made us possible.\footnote{It is like pretending to explain the birth and development of a tree by arguing that the seed from which it grew was produced by the tree itself.} 

To sum up, neither quantum fluctuations, standard unitary evolution, measurements nor decoherence are able to explain the breakdown of the homogeneity and isotropy of the quantum state after inflation. It seems that, in the absence of an additional element, the standard approach is simply unable to accommodate the breakdown of the symmetry. Thus far, we have specified in detail the challenge to be met by any attempt to account for the classicalization of the early universe. In particular, one must account for the transition of the initial, completely homogeneous and isotropic state, into one with inhomogeneities and anisotropies. Moreover, one must show that the resulting inhomogeneous and anisotropic state can, in fact, be described in classical terms. We are, finally, in a position to present and assess the validity of the arguments defended in \cite{ashtekar}. Below, we will argue that all of the attempts considered fail because, either they rely upon unjustified, implicit assumptions regarding the breakdown of homogeneity and isotropy, or they contain the sorts of circular causal explanations mentioned above.

\section{Three paths to classical behavior: the usual suspects}
\label{Ash}

In this section, we describe the three paths to classical behavior in cosmology considered by \cite{ashtekar}: fading of non-commutativity, quantum squeezing, and the transition from quantum states to classical distributions. It is worth noting that similar proposals have been put forward in multiple publications, including \cite{C1,C2,C3,C4,C5,C6,C7,C8,C9}, and have been severely criticized in discussion such as \cite{Shortcomings,D1,D2,Interpretations}. The work in \cite{ashtekar} could then be interpreted as an attempt to overcome those criticisms by strengthening the standard arguments. In what follows we discuss this new proposal, pointing out its deficiencies.

\subsection{Fading of non-commutativity}

It is well known that the failure of observables to commute is a typical marker of quantum behavior. As a result, it might be argued that, if in a certain situation, this non-commutativity becomes negligible (in an appropriate sense), then the system in question might be thought of as behaving classically. In \cite{ashtekar} it is further argued that a fading of non-commutativity could be detected by comparing the expectation values for commutators and anticommutators. They conclude that, if in the cosmological context, the expectation value of the commutator of the canonical operators $\left(\hat{\varphi}(\eta), \hat{\pi}(\eta)\right)$ becomes small compared to the corresponding anticommutator, then one can take this as a sign of the fading of non-commutativity---and hence of the emergence of classical behavior. 

In favor for this strategy, they advance two considerations. First, that for homogeneous and isotropic states, the expectation value of the anticommutator is equal to the expectation value of the corresponding classical observables in a natural distribution on the classical phase space. Therefore, the ratio of the two expectation values can be taken to measure the `importance of the quantum aspects of the system relative to its classical aspects'. Second, that for quantum mechanical systems whose configuration space is a manifold, one naturally associates configuration observables with functions and momentum observables with vector fields. If so, although the classical and quantum commutation relation are of course different, it turns out that the classical and quantum algebras have the same anticommutation relations. As a result, they propose to take the ratio between the expectation values for commutators and anticommutators as a measure of quantum behavior.

Before exploring the soundness of this whole strategy for classicalization, we briefly review the way in which, according to \cite{ashtekar}, in the cosmological context of interest, the expectation value of the commutator of the canonical operators$\left(\hat{\varphi}(\eta), \hat{\pi}(\eta)\right)$ in the Bunch-Davies vacuum becomes small relative to that of the anticommutator. To show this, they consider a FLRW space-time
\begin{equation}
\textbf{g}_{ab}dx^{a}dx^{b} = a(\eta)^{2}(-d\eta^{2} + \vec{dx}^{2}),
\end{equation} 
with $\eta$ the conformal time, related to cosmic time (i.e., proper time associated with comoving observers) $t$ via $a(\eta)d\eta = dt$. For such a scenario, it can be shown that the expectation values in the Bunch-Davies vacuum for the commutator and anticommutator of the single mode operators $(\hat{\varphi}_{\vec{k}}(\eta),\hat{\pi}_{\vec{k'}}(\eta))$ are given by
\begin{gather}
\langle [\hat{\varphi}_{\vec{k}}(\eta),\hat{\pi}_{\vec{k'}}(\eta)]\rangle = i\hbar\delta_{\vec{k},\vec{-k'}}, \\
\langle [\hat{\varphi}_{\vec{k}}(\eta),\hat{\pi}_{\vec{k'}}(\eta)]_{+}\rangle = -2\hbar a^{2}(\eta)Re(e_{k}(\eta)\overline{e}'_{k'}(\eta))\delta_{\vec{k},\vec{-k'}},
\end{gather}
with the functions $e_{k}(\eta)$ a normalized basis of modes satisfying the appropriate Klein-Gordon equation. The absolute value of the ratio of these expectation values is therefore
\begin{equation}
\lvert R_{\varphi,\pi} (\eta) \rvert = \frac{1}{\lvert 2a^{2}(\eta)Re(e_{k}(\eta)\overline{e}'_{k}(\eta)) \rvert},
\end{equation}
where we have considered the only possible case of interest, i.e. $\vec{k}= - \vec{k}'$.

For the case of de Sitter space-times\footnote{In which case we have 
\begin{equation*}
 e_{k}(\eta) = (\frac{1}{a(\eta)} + \frac{iH}{k})\frac{e^{-k\eta}}{\sqrt{2k}},
\end{equation*}
\noindent with $H$ the Hubble constant, so the ratio becomes $\lvert R_{\varphi,\pi} (\eta) \rvert =\frac{k}{Ha(\eta)}.$}
this ratio does indeed decrease, doing so exponentially with the number of e-folds after the mode has crossed the Hubble horizon (see \cite[section III.B]{ashtekar}). However, the authors point out that the ratio does not decrease with $\eta$ in all possible contexts of interest---for example, it does not vanish in the case of the radiation filled universe. As a result, they conclude that the fading is not really a robust criterion for the emergence of classical behavior, as non-commutativity need not fade even when the system does seem to behave classically. That is, they conclude that the fading of non-commutativity is not a necessary condition for classicality. Still, they believe that, at least in certain scenarios, it does signal the emergence of classical behavior.

We just saw that, according to \cite{ashtekar}, the fading of non-commutativity is not a necessary condition for classicality. Now we show that it also isn't a sufficient condition for classical behavior; that is, that the fact that a certain commutator vanishes in a certain situation, does not imply that the situation can be modeled classically. To see this, consider a pair of spin $1/2$ particles in a singlet state. Of course, the components of the spin operator of particle 1 commute with those of particle 2, but that clearly does not imply that we might regard the situation as classical regarding, for instance, the values of those two spins and their correlations. In particular, as has been conclusively shown in \cite{Bell1964} and (even more explicitly, for the case of 3 entangled particles) in \cite{GHZ}, it is simply untenable to assume that, before any measurement is involved, the spins have well-defined values. It was already clear from \cite{ashtekar} that the fading of non-commutativity is not necessary for classicality; now we see that it also isn't sufficient. It seems that the sensible thing to conclude is that there is, in fact, no relation between the fading of non-commutativity, at least as characterized by \cite{ashtekar}, and classicality.

A further problem with the proposed criterion for classicality is that it can be shown to depend on unacceptable arbitrary choices. To see this in a simple example, consider a free particle of mass $m$, which at $t=0$ is in a minimum uncertainty wave function centered at $ x = x_0 $ and $p=p_0$. Now, for the expectation value of the commutator we of course have $ \langle \psi | [\hat X,\hat P] |\psi \rangle =i \hbar$. As for the anticommutator, it can be computed to give
\begin{equation} 
 \langle \psi | \lbrace \hat X,\hat P \rbrace |\psi \rangle = \frac {2t}{m} ( p_0^2 +\sigma^2 ) + 2x_0 p_0 
\end{equation} 
with $ \sigma $ the uncertainty in momentum (which for a free particle is constant in time). The point is that, at any time, we can change the value of the anticommutator---and, in particular, make it vanish---by a change of frame or origin (the expectation value of the commutator will, of course, not change). This means that, according to this criterion, the classicality (or lack thereof) of a system would depend on those choices, which seems utterly unsatisfactory. Yet another complication we would like to mention is that the arguments in favor of the proposal depend on identifying expectation values with possessed values. However, as we explained above, within the standard interpretation, such an identification only obtains when measurements are involved. Therefore, in order to make any sense, the proposal seems to require for measurements to take place in the early universe.

As we explained in section \ref{ClassCos}, any successful explanation of the quantum-to-classical transition in cosmology must account for the breakdown of the homogeneity and isotropy of the quantum state after inflation. It is easy to see that the fading of non-commutativity does not fare well in this regard. In the whole discussion in \cite{ashtekar} regarding this strategy, the underlying quantum state is always assumed to be the Bunch-Davies vacuum. Therefore, even if the fading of non-commutativity would lead to some sort of classicalization, as we proved in section \ref{ClassCos}, the resulting classical state would necessarily share the homogeneity and isotropy with the quantum state. Going back to the example of the singlet, to think that the fading of non-commutativity breaks the symmetry of the Bunch-Davies vacuum, would be like arguing that the commutativity between $ \vec {S_1} $ and $ \vec {S_2}$ breaks the rotational symmetry along the axis joining the two particles. Of course, a measurement of the spins will, in general, break the symmetry and allow a slightly more classical description, in the sense that each particle will have a definite spin orientation, but in the absence of such an external influence, i.e., that resulting from the act of measurement, the symmetry persists.

In sum, regarding the whole idea of associating classicality with commutativity, the fact is that, even in the absence of non-commutativity, there are many other features that make a situation profoundly quantum mechanical. One such feature is entanglement, which is responsible, not only for establishing important correlations, but also for codifying the symmetry of a state. For instance, in the singlet case discussed above, entanglement is responsible for the correlation between the spins of the particles, as well as for the rotational symmetry of the state. In the case of the Bunch-Davies vacuum, something completely analogous occurs. The entanglement present in the state is fundamental in codifying essential properties of the state, including its homogeneity and isotropy, \cite{Wald}.\footnote{In fact, it is easy to show that, for any finite collection of points $\lbrace x_1 .... x_n \rbrace $ on a constant (cosmological) time hypersurface, and for any transformation $L$, corresponding to a rotation or a translation acting on such set of points (operations that are well-defined because of the fact that the corresponding space-like hypersurfaces are flat), all the $n$-point functions involving, say, field $\hat \phi $ or conjugate momentum operators $ \hat \pi $, denoted collectively here as $ \hat \chi $, evaluated in the Bunch-Davies vacuum, are invariant. Namely $ {}_{BD} \langle 0| \hat \chi (x_1) \hat \chi (x_2) ......\hat \chi (x_n) |0\rangle_{BD} ={}_{BD} \langle 0| \hat \chi ( L x_1) \hat \chi (L x_2) ......\hat \chi (L x_n) |0 \rangle_{BD} $. That is, the nature of the quantum mechanical correlations present in the state of interest are precisely of the type that ensures the symmetries of homogeneity and isotropy of the state, so they are, in that sense, analogous to those that occur in the singlet state of a pair of spin-$1/2$ particles in a singlet.}

\subsection{Quantum squeezing}

The phenomenon of quantum squeezing is often employed in arguments for the emergence of classical behavior in the early universe. The idea is that the fact that the uncertainty of the field gets highly squeezed during inflation, and remains so at late times, can be taken as a sign of classicality. In \cite{ashtekar} it is further argued that squeezing can be traced back to geometrical structures in classical phase space and that inflation is not essential for its occurrence. 

Before examining the alleged relation between squeezing and classicality, let us illustrate the issue by reproducing the calculations on the subject, in the context of a de Sitter space-time (see \cite[section IV.A]{ashtekar}). We start by recalling that a quantum system is said to be in a squeezed state when the uncertainty in its canonical variables is not ``evenly distributed''. Now, in the case of de Sitter, the uncertainties for the canonical operators in the Bunch-Davies vacuum are given by
\begin{gather}
\lvert \Delta \hat{\varphi}_{\vec{k}} (\eta) \rvert^2 = \frac{\hbar}{2k}(\frac{1}{a^{2}(\eta)} + \frac{H^2}{k^2})\\
\lvert \Delta \hat{\pi}_{\vec{k}} (\eta) \rvert^2 = \frac{k\hbar}{2}a^{2}(\eta).
\end{gather}
If we focus on modes deep within the Hubble radius at early times, for which $\frac{k}{a(\eta)}\gg H$, we have
\begin{gather}
\lvert \Delta \hat{\varphi}_{\vec{k}} \rvert^2 \approx \frac{\hbar}{2k}\\
\lvert \Delta \hat{\pi}_{\vec{k}} \rvert^2 = \frac{k\hbar}{2},
\end{gather}
where we have, following the convention adopted in \cite{ashtekar}, set $a=1$ at a cosmic time $t=0$. However, after leaving the Hubble horizon, i.e. when $\frac{k}{a(\eta)}\ll H$, for those modes we have
\begin{gather}
\lvert \Delta \hat{\varphi}_{\vec{k}} \rvert^2 \approx \frac{\hbar H^{2}}{2k^{3}}\\
\lvert \Delta \hat{\pi}_{\vec{k}} \rvert^2 = \frac{k\hbar}{2} a^{2}(\eta).
\end{gather}
We see that the uncertainty in the field at late times is very small, compared to early times. This is the phenomenon of squeezing during inflation.

It seems clear that, at least in the sense described above, squeezing indeed occurs during inflation. What we are interested in exploring here is whether such a phenomenon implies that the situation under consideration can be successfully approximated by a classical description. We start by pointing out that in \cite{ashtekar} there are no actual arguments presented in favor of quantum squeezing as a sign of the emergence of classical behavior. The authors, instead, seem to have taken this as a given, focusing only on bringing to the fore how and why squeezing occurs in the cosmological setting. Needless to say, a squeezed state is a quantum state and any talk of it being a sign of classical behavior needs explicit argumentation. 

Presumably, the intuition behind the association of squeezing and classicality stems from the observation that quantum uncertainties are a unmistakably quantum characteristic of a physical system. Therefore, it is contended that, if the uncertainty in a quantum state becomes small, then the state itself becomes `less quantum-like'. There are, however, serious complications with this sort of reasoning. 

To begin with, the fact that a state is squeezed, in no way means that it will not display any of the typical quantum features, such as indeterminism (in agreement with the Born Rule), entanglement, tunneling, being affected by measurements, etc. In fact, squeezed states are very well-known to display acute quantum behavior \cite{Shortcomings,beck,loudon}. This has been exploited by experimentalists working in quantum optics who use squeezed states of light to reduce the photon counting noise in optical high-precision measurements, to calibrate the quantum efficiency of photo-electric photo detectors, or for entanglement-based quantum key distribution.

Moreover, it seems that if one decides to take squeezing as a sign of classicality, that is, if one chooses to regard states with a narrow uncertainty in some property as behaving classically, then one is bound to take states with large uncertainties as very quantum. The issue, of course, is that, just because a certain variable has become squeezed, does not entail that other variables will too. In fact, the contrary is the case: the conjugate of a squeezed variable will get stretched (and often the product of the uncertainties will not even saturate the Heisenberg relation). It seems, then, problematic to take the squeezing of one observable as a sign of classicality; why should we consider a system which is highly localized in its position, but with a highly unlocalized momentum as somehow classical?

As we argued in section \ref{Class}, classicality can only be predicated of a state, relative to a list of observables and associated accuracies. Then, what one could say about, for instance, a particle squeezed in position, is that its state exhibits classical behavior with regards to its position, and only if the uncertainty is small compared to the accuracy of the position measuring apparatus at hand. Moreover, since the squeezing in position implies a stretching in momentum, then all the possible observables of the form $F(\hat{x},\hat{p})$ would exhibit non-classical behavior, due to their dependence on $\hat{p}$. Since the family of operators which only depend on $\hat{x}$ is extremely small in comparison to the set of observables with the general form $F(\hat{x},\hat{p})$, this leaves us with an extremely narrow scope for the relation between squeezing and classicality. 

Finally, as emphasized in \cite{Shortcomings}, and explicitly acknowledged in \cite{ashtekar}, the issue of whether or not the state, at any particular time, is squeezed or not, depends on the quantum operators one chooses to consider in the description of the quantum field. In order to address this rather serious obstacle, \cite{ashtekar} proposes to justify their preferred choice of operators on considerations of what we can in fact measure. As we explained above, this dependence of the explanation of the emergence of classicality on measurements performed by us is simply untenable, as it relies on a circular logic. Again, what we are trying to explain are the primordial conditions that, eventually, led to our existence. It should be clear that it is absurd to allow for considerations regarding what we, humans, can or cannot in fact measure, to play a role in such an explanation.

How does squeezing fare regarding the breakdown of the homogeneity and isotropy of the state? Very poorly indeed. To begin with, it is easy to see that squeezing is perfectly compatible with the symmetry. That is, that the presence of squeezing by itself does not imply a breakdown of homogeneity and isotropy . Moreover, as with the fading of non-commutativity, the whole discussion in \cite{ashtekar} regarding squeezing assumes that the quantum state is the symmetric Bunch-Davies vacuum. We conclude that, even if squeezing were to lead to classicality, the resulting classical state would be homogeneous and isotropic, so the strategy would fail its task to explain the emergence of structure.

All of these considerations prompt us to conclude that there is no sense in which squeezing, in and of itself, entails or is a feature of classicality, as required by the problem at hand. 
 
\subsection{From quantum states to classical distributions}

As we explained above, the standard procedure in contemporary cosmology is to replace the quantum state at the end of inflation with an appropriate classical state---with all subsequent analysis being classical. The question is whether such a procedure can, in fact, be justified from first principles. With this question in mind, in \cite{ashtekar} it is shown that one can associate with the quantum vacuum a distribution function over phase space, such that the quantum expectation value of any Weyl-ordered quantum operator in the vacuum, exactly equals the corresponding classical expectation value. This is taken as a ``clear-cut justification for the procedure used in the early universe literature.'' It is further noted that the proved result is not tied to inflation and that, as long as the perturbations are assumed to be linear, it is valid on any globally hyperbolic space-time. This they take as an indication that non-linear effects, such as mode-mode coupling and decoherence, although quite interesting, are not essential for the emergence of classicality in cosmology.

What are we to make of these assertions? As we just saw, the authors take the equality of expectation values as complete justification for the replacement of the quantum state by the corresponding classical distribution. However, things are not that simple. The problem is that the fact that the expectation values coincide numerically, does not imply that the physical situations they represent are equal. In particular, it is very important to keep in mind that, in general, a quantum expectation value does not represent a possessed property of the system under consideration, it only codifies how the system will behave \emph{when measured}.\footnote{It is interesting to point out that a very similar mistake is commonly made in arguments defending the use of decoherence to explain the absence of macroscopic interference (see \cite[section 2]{LessDeco}).} Therefore, to simply assume that, because the expectation values coincide, one can replace the quantum state by a classical distribution, is in effect to implicitly assume that the quantum vacuum has somehow morphed itself into one of the members of the corresponding classical ensemble. But isn't that just begging the question? Rather than finding a dynamical, or otherwise non-ad hoc, way to justify the quantum-to-classical transition, the strategy simply presupposes that this transition has somehow taken place.

To make these considerations more transparent, let us focus again on the much simpler case of an harmonic oscillator in one dimension (with the center of the potential at $X=0$). If we take the system to be in the ground state (with uncertainties in position and momentum $ \Delta X$ and $\Delta P$), we can construct a suitable distribution function over phase space, such that the quantum and classical expectation values exactly coincide. Does that mean that we might regard a single harmonic oscillator in its ground state as equivalent to an ensemble of points in phase space, with the corresponding distribution? Of course not. It is only if we consider an ensemble of harmonic oscillators in the ground state, and we subject all of them to a measurement of position with a precision higher than $ \Delta X$, that we would have an ensemble of systems with relatively well-defined positions, distributed around the origin with a statistical dispersion of order $ \Delta X$. The crucial point is that, each of the elements of the ensemble, would have undergone a change in its state as a result of the measurement. It is, then, only through measurements that the quantum expectation values get connected with the statistical characteristics of the ensemble that results from the measurement performed on all the systems.
 
Regarding the breakdown of the homogeneity and isotropy, it must be noted that this last strategy considered in \cite{ashtekar} is the only one in which an attempt is made to deal with the passing from a symmetric to a non-symmetric situation. In order to assess its success, consider the reflection symmetry $ X \to -X$ of the harmonic oscillator. Such a symmetry is present, both at the dynamics of the system (i.e., its Hamiltonian), and in the ground state. Now, can a substitution of the ground state by the corresponding distribution over phase space---made on paper by some theoretical physicist, with not actual physical counterpart---be used to argue that the actual physical situation has lost its symmetry? Of course not, unless a measurement is involved. In exactly the same way, a mere substitution of the Bunch-Davies vacuum by a classical distribution function with the same expectation values, does not constitute a breakdown of the symmetry. It is often argued that this prevalence of the symmetry must be read as indicating that the whole ensemble retains full homogeneity and isotropy, but that each member is not required to be symmetric. However, this answer only works if, again, one implicitly, but illicitly, assumes that the system stops being the Bunch-Davies vacuum and transforms into one of the members of the ensemble. Of course, some sort of measurement would be able to achieve this but, as we have explained above, measurements cannot play any role in this sort of explanation.
 
\section{Conclusions}
\label{Conc}

The standard story for the emergence of classicality in the early universe asserts that quantum fluctuations, or uncertainties, associated with the completely homogeneous and isotropic Bunch-Davies vacuum, constitute the primordial seeds of all cosmic structure. Based on these ideas, what is usually done in practice is that, towards the end of inflation, the symmetric quantum vacuum is replaced by an appropriate distribution function over classical phase space. A vital question is whether such a standard procedure is indeed justified, i.e., if it follows from first principles, or if it depends upon some additional, possibly invalid, assumptions. 

Over the years, cosmologists have tried to account for this transition by different arguments and methods. A recent work by Ashtekar et al., \cite{ashtekar}, explores three different ways in which classical behavior has been argued to emerge in the early universe: i) the fading of the importance of quantum non-commutativity, ii) the phenomenon of quantum squeezing, and iii) the ability to approximate the quantum state by a distribution function on the classical phase space. They conclude that these notions provide a quite robust explanation for the emergence of classicality in cosmology.

In this work, we dispute the assertion that these or similar accounts constitute a valid justification for the transition to a classical description. We claim that they fail because, they either rely upon unjustified, implicit assumptions, or they contain some kind of invalid, circular logic. Moreover, we point out that any successful effort to explain the emergence of classicality in cosmology must account for the breakdown of homogeneity and isotropy, and we show that none of the considered proposals is able to do so. We conclude that the proposals considered are unable to explain the emergence of classical behavior in the early universe.

More generally, we pointed out that when attempting to explain the emergence of structure in cosmology, one is actually dealing with two issues which are, in principle, distinct and independent, but that, in the context at hand, appear closely connected: A) classicalization and B) breakdown of homogeneity and isotropy. Now, in principle, one might attempt to deal with them in different orders: first A and then B or vice versa. As we have seen, most of the existing attempts to address the problem at hand, including two of the three considered in \cite{ashtekar} (with the last one suffering from circularity in its explanatory power) focus on A without even considering B. It should be clear that any attempt to follow such path is essentially doomed to fail. This is because one would be trying to account for the breakdown of homogeneity and isotropy in classical terms, but in a manner that has no quantum mechanical counterpart. As we argued in section \ref{ClassCos}, that is inconsistent with the basic assumption that the quantum description is more fundamental than the classical one---and thus that any classical characterization supervenes on a quantum one. 
 
Finally, it seems clear that when attempting to deal with A without considering B, one will be confronted with the cosmological version of the measurement problem. In this regard, one should heed the lessons from Bell \cite{Bell81} and others, and, in particular, the fundamental result of \cite{Tim}, showing the intrinsic inconsistency of simultaneously holding the following three claims: 1) the physical description given by the quantum state is complete; 2) quantum evolution is always unitary; 3) measurements always yield definite results. The measurement problem is the elephant in the room in many situations of physical interest, and ignoring it might lead one astray.

\section{Reply to `Note added in response to arXiv:2009.09999'}
\label{reply}

\textbf{This is our reply to a note added in arXiv:2004.10684 by Ashtekar, Corichi and Kesavan (henceforth ACK) in response to this paper. We have inserted our comments in bold into their note.}\\

Javier Berjon, Elias Okon, and Daniel Sudarsky (referred to as BOS in what follows) submitted an article to a journal. Following their general policy, Editors asked us to comment on this paper because it ``appears to be critical of some aspects of reference 1 of the manuscript, which you coauthored'' (namely, this paper). We responded to the journal with the following comments. Since the arXiv submission has remained unchanged, we are now putting our response in the public domain.\\

1. We are perplexed by this paper. Perplexed, because while the wording in BOS suggests that our paper ``Emergence of classical behavior in the early Universe'', Physical Review D 102, 023512 (2020) (referred to as ACK below) is the center of their criticism, the critical statements refer to claims in other, older works, not to results in our paper. Further, while, starting already in the abstract, the BOS paper criticizes the literature for ``lack of clarity'' because of reliance on ``unjustified and implicit assumptions'', we encountered these very problems in the BOS paper! In what follows we explain these two points in some detail.\\

\textbf{We must say that we find quite surprising the claim that the criticisms in our work do not apply to the arguments in ACK. As we clearly state in the abstract and introduction, our work critically explores the validity of different efforts at explaining the emergence of classical behavior in cosmology. ACK, for their part, exactly sets out to explore some of these efforts and, for the most part, finds them successful---making it a clear and fair target of our critique. At any rate, below we will explain in full detail why our arguments indeed fully affect the results and conclusions in ACK.}

\textbf{As for the charge of ``lack of clarity'' in our work, we believe that it has more to do with a careless reading of our manuscript, than with actual conceptual mistakes on our part. In any case, below we will clarify any possible misunderstanding on the part of ACK regarding our arguments.}\\

2. Already the abstract of the ACK paper emphasizes that we \emph{investigate} three issues that have been discussed in the context of inflation, and the concluding part of the paper emphasizes that the goal was to discuss these issues \emph{from a mathematical physics perspective}. We also emphasize in this part that issues such as non-linearities, mode-mode couplings, decoherence, quantum discord and quantum measurement theory \emph{are important and will have to be addressed in a more complete discussion}. In contradistinction to what the wording used by BOS suggests, we never say that our results provide a complete account. Nowhere did we claim or suggest that any of the three criteria we discuss implies that the state becomes inhomogeneous and anisotropic, or lead to what BOS refer to as ``classicalization''. In fact, already in our Introduction, we emphasize that in our analysis ``there is no quantum to classical transition'', whence we ``will not need to enter a discussion of issues that arise when the focus is on transition''. We also say ``Rather, our emphasis is on the ``emergence'' of classical behavior ... in the mathematical description of cosmological perturbations'' and our notion of emergence refers to the dynamics of specific sub-classes of observables. Therefore, the main thrust of the negative comments in the BOS paper has very little to do with what we actually discuss and prove.\\

\textbf{As we said above, we find the claim that our critique has very little to do with ACK's work astounding. To begin with, the title of the paper is ``Emergence of classical behavior in the early universe'', which already suggests that the paper engages with explanation regarding, well, the emergence of classical behavior in the early universe---which is exactly the target of our critique. Moreover, it is true that ACK emphasizes that issues such as mode-mode coupling, decoherence and quantum measurement theory are important, but they also claim that they ``are not essential for the emergence of classical behavior in the early universe'', or that ``classical behavior emerges without them in the early universe.'' In other words, they explicitly claim that the effects they do explore are sufficient to explain the emergence of classical behavior. As a result, they state, ``Our emphasis will be on isolating the simplest mechanisms that can lead to classical behavior in the early universe'', and they conclude: ``Our discussion brought out the precise sense in which classical behavior emerges in the early universe already in the simplest context --that of the quantum theory of linear cosmological perturbations.'' It seems quite clear that, in spite of what is stated above, ACK's claims go well beyond a ``pure mathematical analysis'' and do engage precisely with the arguments we critically explore in our work.}

\textbf{Regarding the claim that our analysis refers to instances of ``classicalization'' and ``quantum to classical transition,'' while ACK only considers the ``emergence of classical behavior'', we point out that, right at the beginning of our work, we emphasize that the notions of ``classicalization'' and ``quantum to classical transition'' should not be interpreted literally. In other words, we assert, right from the start, that what we explore is the notion of ``emergence of classical behavior.'' In any case, we do not mind if a reader chooses to substitute any instance of the phrase ``quantum to classical transition'' with ``emergence of classical behavior'', as that would not change at all the critical content of our paper.}\\

3. What we actually do in ACK is to carefully examine the three criteria that have been widely used in the literature in the discussion of the emergence of semi-classical behavior and clarify the precise sense in which the statements in the literature hold---sometimes upon appropriate modifications---and the sense in which some of the mathematical arguments are incomplete, or even incorrect. So the emphasis is on a \emph{critical mathematical investigation}. In particular, some of the literature had implicitly assumed that the three criteria are different facets of the same underlying phenomenon and we show that this is not the case.

It is only the Section 4 of the BOS paper that directly addresses the results of ACK.

\noindent (i) They agree with us that the first criterion---fading of non-commutativity---has some important limitations.\\

\textbf{It is true that ACK acknowledges certain limitations of the fading of non-commutativity as a criterion of classicality. In particular, they show that the fading of non-commutativity is not a necessary condition for classicality. However, they also claim that, at least in certain scenarios, non-commutativity does signal the emergence of classical behavior. That, of course, is very different from our own conclusion on the matter, which is that non-commutativity is neither necessary nor sufficient for classically, i.e., that there is, in fact, no relation between the fading of non-commutativity, at least as characterized by ACK, and classicality.}\\

\noindent (ii) The BOS summary of what motivated previous works to use squeezing as a signature of the emergence of classical behavior is the same as ours. But then they say that in the ACK paper there are ``no actual arguments presented in favor of quantum squeezing as a sign of the emergence of classical behavior''. This is incorrect. ACK have emphasized that their use of the term ``classical behavior'' refers to a set of observables and their uncertainties in a given state. (BOS also state this as their own viewpoint on page 8). Classical behavior emerges for the field operators in which the uncertainty is squeezed, and ACK also provide a physical understanding as to why it continues to remain squeezed as time evolves in spite of the fact that the uncertainty in field momenta is large. The main point of this Section in ACK is to trace back the phenomenon of squeezing to geometrical structures on the classical phase space. BOS have no comment on that.\\

\textbf{We stand by the claim that, in the ACK paper, there are ``no actual arguments presented in favor of quantum squeezing as a sign of the emergence of classical behavior''. Of course, ACK does point out that ``The phenomenon of quantum squeezing is often used to argue that classical behavior naturally emerges during inflation...'' but that is not an argument, it is just an assertion. The question, then, is whether it is reasonable to use squeezing as a signature of the emergence of classicality. In this regard, we end up concluding that there is no sense in which squeezing, in and of itself, entails or is a feature of the emergence of classicality.}\\

\noindent (iii) The third criterion is perhaps the most important one ``in practice'' because as ACK explain---and BOS reiterate---in most calculations one generally replaces the quantum state of perturbations by a distribution function on their classical phase space. Because a state is completely determined by the expectation values of all observables both in classical and quantum mechanics, \emph{time evolutions} of expectation values provide a natural avenue to critically investigate the validity of this procedure. Using the Bargmann representation for quantum states of perturbations, we obtained a sharp mathematical result on when the procedure is justified, and when it is not. Our results are more general than what was known in the cosmology literature before. We find that the class of observables is surprisingly large, much larger than products of just the field operators, or, just their momenta, that were typically considered. For the larger class of observables, then, the mathematical procedure used in the literature to time-evolve these observables is justified. The BOS paper has no comment on this main result. Their criticism is that this procedure does not explain the breakdown of homogeneity and isotropy. ACK never said that it does. Thus, the criticisms in the BOS Sections 4 and 5 have almost nothing to do with the ACK results.\\

\textbf{It is true that we explain that it is standard procedure to substitute the quantum state after inflation by a distribution function on classical phase space. The important question we are interested in, and which ACK poses explicitly, is whether such a procedure can be justified from first principles. The answer given by ACK is that their analysis ``provides a clearcut justification for the procedure used in the early universe literature''. It is this statement which we critically explore in our paper. Our conclusion, in contradistinction to what ACK suggests, is that such a procedure, while widely used, cannot really be justified from first principles within the standard framework.}

\textbf{One can see explicitly the level of confusion underlying ACK argumentation in this point by noting that the same arguments might be applied to the ground state of the harmonic oscillator. If so, according to ACK, a harmonic oscillator in its ground state might be regarded as a distribution of classical states, with rather well-defined values of position, momentum or some other choice for canonical variable. Were one to do so, one would have to conclude that the distribution one is talking about is formed by sharply peaked states in, say, position. And, if so, all the states of the distribution would have energies that are much higher than that of the ground states, so they would not be physically equivalent to the state one started with. ACK evades these and related concerns simply by using the word ``equivalent'' in a rather loose manner, which only serves to mislead the reader and contribute to the existing confusion on the subject among certain sectors of the community.}\\

4. In the BOS paper, we found several instances of ``lack of clarity'' because of reliance on ``unjustified and implicit assumptions''.

Page 6: ``It is important to point out that, in order to apply our proposed criterion for classicality, it is not necessary for actual measurements to ever take place; that is, a system can be deemed to have classicalized, even if there are no observers around to verify it. Still the notion of measurements remains central, as it does in the standard quantum interpretation.''
Central, for what? Presumably not for ``classicalization'', since that is what the previous sentence states. But isn't the entire BOS focus on what they call ``classicalization''?\\

\textbf{From the quoted paragraph, it is clear that what is meant is that the notion of measurement is central for the definition of classicality. The point is that, even though, in order to define classicality, we consider how a system would behave if measured, the system in question does not, in fact, have to be measured in order to fulfill the criterion. That is, a system does not have to actually be measured for it to become susceptible to an approximated classical description. We conclude that there isn't any lack of clarity in this part of our text, but rather a very careless reading on the part of ACK}\\

Page 7: ``However, it becomes untenable as soon as one intends to employ the theory in more complex scenarios, such as the early universe, in which there are no observers around to measure, ...''
But didn't BOS say on page 6 that criterion for ``classicality'' does not need an observer?\\

\textbf{Contrary to what is stated above, and as should be quite clear from context, the ``it'' in the quoted passage does not refer to our criterion for classicality, but to the standard interpretation of quantum mechanics. There is, then, no contradiction between such a passage and what we say in page 6.}\\

Page 8: ``In particular, we saw that classicalization can only be defined relative to a set of observables, with associated uncertainties, and that a given state of the system might be describable in classical terms when focusing on a certain feature, but not when focusing on others.''

\noindent This is a reiteration of the ACK viewpoint. But then we are puzzled by the ``problem'' discussed on page 18:

\noindent ``The problem, of course, is that, just because a certain variable has become squeezed, does not entail that other variables will too.''

\noindent Why is this a problem? After all, criterion BOS fully accept on page 6, ``there is no absolute sense in which a quantum system might be said to classicalize this can only be asserted relative to the cited additional information.''\\

\textbf{As it is clearly explained in the paragraph immediately after the one from which the above quote is extracted, the real problem we point out is that the fact that the conjugate of a squeezed variable necessarily gets stretched, implies that the association between squeezing and classicality has an extremely narrow scope.}\\

Page 9: ``Now, in the same way that a breakdown of homogeneity and isotropy does not imply classicality, classicality can occur in the absence of a breakdown of the symmetry. Nevertheless, what we want to point out is that, classicalization by itself, cannot erase or eliminate a symmetry present at the quantum level and, in particular, it cannot break homogeneity and isotropy.''

\noindent ACK never say that emergence of the classical behavior in any of the senses discussed implies breaking of homogeneity and isotropy. So what is the point that BOS are making?\\

\textbf{As we explained above, ACK concludes that the effects they explore are sufficient to explain the emergence of classical behavior in the early universe. We, on the other hand, show that the breakdown of homogeneity and isotropy is a necessary requirement for a successful explanation of such an emergence of classicality. The unavoidable conclusion is that it is actually not the case that the explored effects are, as argued by ACK, able to deliver the desired explanation.}

\textbf{It is also important to point out that ACK fully embraces the standard position according to which ``the origin of the large scale structure of the universe is traced back to quintessential quantum fluctuations that cannot be switched off even in principle''. That, of course, is a key element of the position we are criticizing. We must say we find it rather disingenuous to try to argue, as ACK is doing at this point, that the discussion of their paper is separated from that very issue.}\\

Page 10: ``any successful effort to explain the emergence of classical behavior in the standard cosmological context must necessarily involve a satisfactory account of the breakdown of the homogeneity and isotropy of the quantum state.''

\noindent This seems to contradict their assertion on page 9 that the two are unrelated. Also, this conclusion seems to be too sweeping and therefore lacking in precision/clarity. In fact the quote above from page 8 says that this ``emergence'' can be defined relative to a set of observables and that procedure does not require breakdown of homogeneity and isotropy.\\

\textbf{In the text, we do note that the emergence of classical behavior and the breakdown of the symmetry are logically independent. However we carefully and explicitly show that ``if the quantum state of the system is indeed fully homogeneous and isotropic, as it is taken to be after the first few e-folds of inflation, then, even if a classical description would be available, such a description would also have to be fully homogeneous and isotropic.'' Next, we note that any successful description of the universe must involve the presence of some sort of structure, so we conclude that any successful description of the universe must involve a satisfactory account of the breakdown of the symmetry.}\\

5. \emph{Summary}: The only way we can possibly understand the perplexing BOS submission is through the concern they express in the opening para of their Section [4]: The ACK work could ``then be interpreted as an attempt to overcome'' the criticisms of [8,10,19,24]``by strengthening the standard arguments'' made, e.g. in [2, 6,12-16,18, 23].\footnote{Refs. [8,10,19, 24] refer to older works in inflation, not to what we actually do (see Point 3 above). [10] is on the Continuous Spontaneous Localization idea and [19] is a criticism of that idea; they are not even tangentially related to our results! [8, 24] are co-authored by Sudarsky.} This is a misinterpretation made by BOS, and their entire paper appears to be a reaction in defense of the arguments made in [8,24], based on this misinterpretation.\\

\textbf{It is unfortunate that one must remind people that physics is not simply math. That a clear \emph{physical interpretation} of the symbols employed is essential, particularly if one wants to make substantial claims about the universe. Here, ACK seems to be trying to, as they say, have their cake and eat it too. They claim that they don't have to respond why they assume squeezing or non-commutativity entail some sort of emergence classical behavior, because they are merely `focusing on the math'. However, they do make use of these assumptions to conclude that emergence of classical behavior is obtained from first principles during the post-inflation period in the universe. It seems clear to us that such a position is simply untenable.}

\textbf{In short, ACK's work dots some i's and crosses some t's on arguments that, on much broader and conceptual grounds, have been shown to be unsuitable to account for the physics that needs to be explained. Then, without any solid justification, the claim to have accomplished the latter feat.}

\section*{Acknowledgments}
D.S. acknowledges partial financial support from PAPIIT-UNAM grant IG100120; CONACYT grant 140630; the Foundational Questions Institute (Grant No. FQXi-MGB-1928); the Fetzer Franklin Fund, a donor advised by the Silicon Valley Community Foundation. E.O. acknowledges support from UNAM-PAPIIT grant IN102219.

\bibliographystyle{plain}
\bibliography{bibClass}


\end{document}